\documentclass[aps,prl,twocolumn,reprint,amsmath,amssymb,showpacs,superscriptaddress]{revtex4}
\usepackage{graphicx,color,hyperref}
\begin{document}
\def\be{\begin{equation}}
\def\ee{\end{equation}}
\def\bea{\begin{eqnarray}}
\def\eea{\end{eqnarray}}
\def\bef{\begin{figure}[h!]}
\def\eef{\end{figure}}

\def\a{\alpha}
\def\th{\theta}
\def\o{\omega}
\def\eps{\epsilon}
\def\fr{\frac}
\def\l{\label}
\newcommand{\ra}{\rangle}
\newcommand{\la}{\langle}

\title{Mixing and Ergodicity in Systems with Long-Range Interactions}

\author{Tarc\'isio N. Teles}
\affiliation{Grupo de F\'isica de Feixes, Universidade Federal de Ci\^encias da Sa\'ude de Porto Alegre (UFCSPA), Porto Alegre, RS, Brazil}

\author{Renato Pakter}
\author{Yan Levin}
\email{levin@if.ufrgs.br}
\affiliation{Instituto de F\'isica, Universidade Federal do Rio Grande do Sul (UFRGS), Porto Alegre, RS, Brazil}

\date{\today}

\begin{abstract}
We present a theory of collisionless relaxation in systems with long-range (LR) interactions. Contrary to Lynden-Bell's theory of violent relaxation, which assumes global ergodicity and mixing, we show that the quasi-stationary states (qSS) observed in these systems exhibit broken global ergodicity. We propose that relaxation towards equilibrium occurs through a process of local mixing, where particles spread over energy shells defined by the manifold to which their trajectories are confined, accomplished by halo formation typical of the relaxation in LR systems.  To demonstrate our Local Mixing Core-Halo (LMCH) theory, we study the Hamiltonian Mean Field (HMF) model, a paradigmatic system with long-range interactions. Our theory accurately predicts the particle distribution functions in the qSS observed in molecular dynamics simulations without any adjustable parameters. Additionally, it precisely forecasts the phase transitions observed in the HMF model.
\end{abstract}

\date{\today}
\maketitle


\textcolor{black}{Since the pioneering works of Clausius, Boltzmann, and Gibbs, it has been established that systems with short-range (SR) forces relax to thermodynamic equilibrium~\cite{Campa:2009}. On the other hand, systems with long-range (LR) interactions --  for which the pair potential decays as a power law with an exponent equal to or less than the spatial dimensionality  --  are significantly more complex~\cite{Levin:2013}. LR interaction makes these systems intrinsically non-extensive.   The extensivity can be recovered via the Kac prescription~\cite{Kac:1963}, which scales the strength of the interaction potential with the number of particles. However, even in this ``mean-field" limit, LR systems remain non-additive, exhibiting phenomena like ensemble inequivalence, negative specific heat, and temperature jumps in the microcanonical ensemble~\cite{Barre2001,levon,Thirring,Lapo:2015}.}


\textcolor{black}{From a dynamical perspective, LR systems are characterized by diverging collision times, which can exceed astronomical timescales, as seen in elliptical galaxies~\cite{Pad90}. In the limit of a large number of particles, these systems do not reach thermodynamic equilibrium but instead become trapped in non-equilibrium quasi-stationary states (qSS) whose lifetime diverges with the system size~\cite{Touchette2009,Mukamel}. The ubiquity of qSS in systems ranging from globular star clusters to cold atom systems and magnetically confined plasmas~\cite{Levin:2008,Bouchet2012,Chavanis2000,ruffoquantum,Slama2007} underscores the importance of studying collisionless relaxation.}


For systems with  SR interactions, the state of thermodynamic equilibrium is unique  -- it is  independent of the initial particle distribution, and only depends on the conserved quantities.  On the other hand qSS to which systems with LR interactions evolve depends explicitly on the initial particle distribution. The evolution of the one-particle distribution function \( f(\vec{r}, \vec{p}, t) \) is governed by the Vlasov equation, analogous to the dynamics of an incompressible fluid in a six dimensional phase space~\cite{Braun:1977}. However, solving this equation numerically poses significant challenges, leading most studies of LR systems to rely on Molecular Dynamics (MD) simulations.



A number of statistical theories attempting to explain the nature of qSS have been proposed over the years. In a pioneering work Lynden-Bell (LB) proposed that there is ergodicity and \textcolor{black}{global} mixing of the density levels of the initial distribution function during the dynamical evolution of the system~\cite{Lynden:1967}.  Under these assumptions the final qSS  corresponds to the state that maximizes Boltzmann entropy subject to the conservation of the Casimir invariants of the Vlasov dynamics~\cite{marcianoPRL}. The resulting distribution functions obtained using such approach, however, were found to deviate significantly from the results of MD simulations even for the simplest one-level water-bag initial distribution~\cite{Pakter:2011,Teles:2012,Benetti:2014,Lapo:2021,Ewart,Benetti:2012}.

\begin{figure}[!h]
\includegraphics[scale=1.25,width=8.cm]{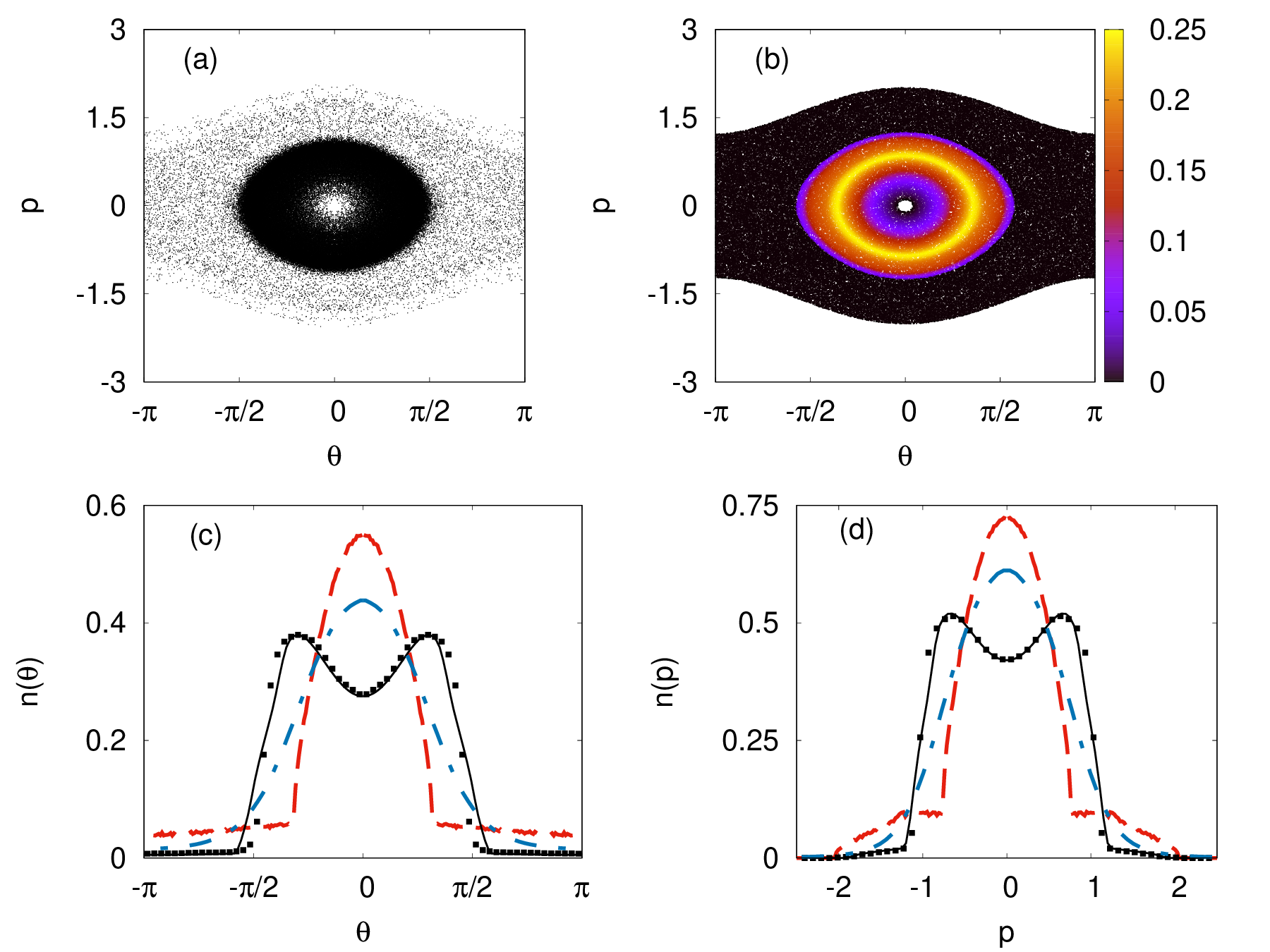}

\caption{\textcolor{black}{Panel (a) shows a snapshot of the particle distribution in the qSS obtained using molecular dynamics (MD) simulations with $N=10^5$ particles. Panel (b) presents the distribution function calculated using the LMCH theory presented here, with the color scale representing its values. Panel (c) depicts the marginal particle distribution in angle, while panel (d) shows the marginal distribution in momentum, both compared to MD simulations (symbols). The initial distribution, given by Eq. (\ref{f0}), has $m_0 = 0.475$ and $u_0 = 0.5$, with the qSS magnetization evolving to $m_{qSS} = 0.63$. The magnetization in the qSS is obtained by averaging \( m(t) \) over a time window after the qSS is reached, yielding \( m_{\mathrm{qSS}} \equiv \bar{m}(t) \). The dot-dashed blue curves represent predictions from LB theory, calculated using the Monte Carlo simulation method described in reference~\cite{entropy}, while the dashed red curves correspond to predictions from CH theory~\cite{telesAPL}. The black solid curves represent the predictions of the LMCH theory presented in this paper and points are the results of MD simulations.
}}
\label{figinitial}
\end{figure}

\textcolor{black}{A different approach introduced in 2008 posits that LR systems relax to a core-halo (CH) distribution with a fully degenerate core~\cite{Levin:2008}. Such distributions arise when ``wave-particle" interactions and Landau damping govern the system's evolution~\cite{Levin:2013}. Numerical observations of core-halo structures trace back to early simulations by Lecar and Feix of self-gravitating one-dimensional systems~\cite{Lecar,Feix}. The boundary of the halo region can be understood through the dynamics of test particles interacting with an oscillating mean-field potential~\cite{Gluckstern:1994}. The parametric resonances responsible for the energy gain of some particles lead to evaporative cooling of the core. The CH theory suggests that evaporation will cease when all the low energy states of the core are fully occupied, up to the maximum density allowed by the incompressible Vlasov dynamics~\cite{Levin:2013}. For an initial water-bag particle distribution, the CH theory predicts the resulting qSS across a wide range of systems~\cite{Levin:2013}. However, both LB and CH theories have recently been shown to fail with more complex continuous initial distributions, particularly when initial particle distributions exhibit population inversion  --- with  high-energy states having more particles than the low-energy states~\cite{entropy}, as illustrated in Fig.~\ref{figinitial}. Under such conditions, the final qSS neither aligns with the LB theory nor reflects the fully degenerate core suggested by the CH theory.}

In this paper, we will present a new theory that is able to quantitatively account for the results of MD simulations for arbitrary initial conditions without any adjustable parameters.
To explain the new theory we will consider the paradigmatic model of a system with long-range interactions the, so called, Hamiltonian Mean Field (HMF) model.  HMF can be considered a dynamical version of the classic XY spin chain, or as  $N$  particles moving on a ring, the dynamics of which is governed by the Hamiltonian
\begin{equation}
H=\sum_{k=1}^N\frac{p_k^2}{2}+\frac{1}{2N}\sum_{k,j=1}^N[1-\cos(\theta_k-\theta_j)],
\label{eq:HMF_Hamiltonian}
\end{equation}
where $\theta_k \in [-\pi, \pi )$ denotes the canonical position coordinate of a particle $k$, and $p_k$ is its conjugate momentum~\cite{Ruffo:1995}. Note that all the particles interact with each other.  The model also corresponds to the lowest order approximation of a periodic 1d gravity with a ``neutralizing"  background~\cite{Levin:2008}.  The Kac $1/N$  prefactor is included to make energy extensive~\cite{Kac:1963}. The particle dynamics is governed by the Hamilton equations of motion:
\begin{equation}
\begin{cases}
\dot{\theta}_k=p_k, \\
\dot{p}_k=-m_x(t)\sin\theta_k+m_y(t)\cos\theta_k,
\end{cases}
\label{eq:eom}
\end{equation}
where $m_x$ and $m_y$ represent the components of the magnetization vector ${\vec m}(t)=(m_x(t), m_y(t))=\frac{1}{N}\sum_{k=1}^N(\cos \theta_k,\sin \theta_k)$. \textcolor{black}{To simplify the presentation, we will focus on symmetric distributions for which \(m_y(t) = 0\) throughout the evolution. Therefore, the dynamics is fully characterized by \(m_x(t)\), which for brevity we shall denote as \(m(t)\). The system's dynamics conserves the total energy which can be written as  \(U = N \langle \varepsilon \rangle - N (1-m^2)/2\), where $\varepsilon(\theta,p,t) = \frac{p^2}{2} + 1 - m(t)\cos \theta$ represents the single particle energy and $\langle \cdot \rangle$ is the average over all the particles.} 

\textcolor{black}{The theory allows us to study any arbitrary initial particle distribution.}  For concreteness, let us 
suppose that initially particles are distributed according to 
\begin{equation}
f_0(\theta,p)=\frac{2}{\pi p_0 \theta_0}\left(\frac{\theta^2}{\theta_0^2}+\frac{p^2}{p_0^2}\right) \Theta\left[1-\left(\frac{\theta^2}{\theta_0^2} + \frac{p^2}{p_0^2}\right) \right], 
\label{f0}
\end{equation}
where $\Theta$ is the Heaviside step function.
This parabolic distribution exhibits  a population inversion  -- more particles are at higher energy than at low.  
The initial energy $u_0$ and initial magnetization $m_0$ per particle can be varied by changing the values of $\theta_0$ and $p_0$. 
Since this distribution is not a stationary solution of the Vlasov equation, the dynamics of particles will follow Eqs.~\ref{eq:eom}, eventually evolving to the final qSS shown in Figure~\ref{figinitial}.
Clearly, the  particle distribution in the qSS is very different from the predictions of LB (\textcolor{black}{blue dot dashed curves}) \textcolor{black}{and CH theories (\textcolor{black}{red dashed curves})}. On the other hand,  the solid black curves show the results of the theory that will be presented in this paper. As can be seen, it captures very well the structure of the qSS, without any adjustable parameters. 

The failure of LB theory to correctly account for the particle distribution in the qSS  can be traced to the 
breakdown of the key assumption of ergodicity and \textcolor{black}{global} mixing on which the theory is based. Similarly the assumption of the CH theory that evaporative cooling will result in a completely degenerate core is also not realized.   In this paper we will, therefore, proceed to relax the LB assumption of global ergodicity to a much less restrictive assumption of local mixing along the isoenergy trajectories
defined by the single-particle energy~\cite{Leoncini:2009}.   

The local phase mixing means that for a fixed initial magnetization \(m_0\), particle from the initial distribution \(f_0(\theta,p)\) that have the same  energy $\epsilon$ will uniformly spread over the isoenergy trajectories 
resulting in a coarse-grained particle distribution given by:
\begin{equation}
\label{coarse}
\bar f_0(\epsilon) = \frac{\iint {\rm d\theta} \, {\rm dp} \, f_0(\theta,p)\delta[\varepsilon_0(\theta,p)-\epsilon]}{\iint {\rm d\theta} \, {\rm dp} \, \delta[\varepsilon_0(\theta,p)-\epsilon]}\,,
\end{equation}
where $\varepsilon_0(\theta,p)=\varepsilon(\theta,p,t=0) $. 
The numerator of Eq. (\ref{coarse}) corresponds to the fraction of particles in \(f_0(\theta,p)\) with energies in the interval \([\epsilon;\epsilon+d\epsilon]\), while the denominator to the total phase space volume for this energy interval. 

The integrals over momentum in Eq.~\ref{coarse} can be evaluated using the well known property of the Dirac delta function  \(\delta[h(z)] = \sum_j \frac{\delta(z - z_j)}{|h'(z_j)|}\), where \(z_j\) is the \(j\)-th root of \(h(z)\). Furthermore, the integral over $\theta$ in the denominator can be performed explicitly in terms of elliptic integral functions, resulting in:
\begin{equation}
\label{coarse2}
\bar f_0(\epsilon) = \frac{\int {\rm d\theta} \, \frac{f_0[\theta,\sqrt{2(\epsilon-1+m_0\cos\theta)}]}{\sqrt{2(\epsilon-1+m_0\cos\theta)}}}{g_0(\epsilon)}\,,
\end{equation}
where,
\begin{equation}
g_0(\epsilon) = \begin{cases} 
      2 \sqrt{|m_0|} K[\kappa_0(\epsilon)] & \text{if } \kappa_0(\epsilon) \leq 1 \\
      2 K[\kappa_0^{-1}(\epsilon)]/\sqrt{|m_0| \kappa_0(\epsilon)} & \text{if } \kappa_0(\epsilon) > 1 
   \end{cases}
\label{g0}
\end{equation}
with \(\kappa_0(\epsilon) = (\epsilon - 1 + |m_0|)/(2|m_0|)\) and \(K[x]\) is the complete elliptic integral of the first kind. 
\textcolor{black}{The suggestion that the  qSS observed in the HMF can be described by the expression  Eq. (\ref{coarse2}) was made previously by the authors of references~\cite{Leoncini:2009} and~\cite{Buyl:2011}.   However, very soon it was shown that the qSS described by such approach were restricted to a very special class of initial distributions that satisfy the, so called, viral condition  -- see in particular the discussion in~\cite{Benetti:2012} and~\cite{Teixeira:2014}. Outside of these initial distributions, the magnetization evolves over time, leading to significant discrepancies between theoretical predictions and simulations. This difficulty was already noted by the authors of reference~\cite{Buyl:2011}, who observed that the magnetization of some qSS predicted by their theory was half the value obtained using MD simulations.  Furthermore, there is an additional problem which has gone largely unnoticed  -- the coarse-grained qSS calculated using Eq. (\ref{coarse2}) does not conserve the energy of the system.  
}

\textcolor{black}{To clearly see this, we consider an initial scenario where all HMF particles are located at two distinct points in the phase space, $(\theta, p) = (0, \pm p_0)$. In this configuration, the total energy of the system is $U_0 = N\varepsilon_0 = N p_0^2/2$, with an initial magnetization of $m_0 = 1$. The resulting coarse-grained distribution function after the particles spread over the isoenergy surface according to Eq. (\ref{coarse2}) should be \( \bar f_0(\epsilon) = g_0(\epsilon)^{-1} \),  the energy of which is calculated to be:
\begin{equation}
\frac{U}{U_0} = 1 - \frac{p_0^2}{4} \left(1 - \frac{E[\kappa_0^{-1}(\varepsilon_0)]}{K[\kappa_0^{-1}(\varepsilon_0)]}\right)^2,
\end{equation}
where $E[x]$ is the complete elliptic integral of the second kind.    Since the second term on the right hand side of the equation is always positive, we see that the energy of the the predicted qSS is lower than the energy of the initial distribution, which is clearly not acceptable for a Hamiltonian system.}


\textcolor{black}{The purpose of the present paper is to show how  the approach above can be combined with the ideas of CH theory to correctly predict the final qSS reached by systems with LR interactions. 
To do this we first observe that  the energy released during the mixing process is carried away by the particles that are excited by the parametric resonances, resulting in evaporative cooling of the core region. As the process of evaporative cooling proceeds, the magnetization changes, until the final qSS is established. There are two different time scales  -- a very fast scale of local mixing over the energy shells (the coarse-graining) -- and a slow timescale on which the coarse-grained distribution function evolves.  Using this insight we can then calculate the qSS  iteratively.  The initial fast mixing is described by Eq. (\ref{coarse2})  with iteration index $i = 0$. The magnetization in the iteration $i+1$, corresponding to the slow timescale evolution of the coarse-grained distribution function can be related to the state of the system in iteration $i$:}
\begin{equation}
\label{eq:map1a}
m_{i+1} = \iint {\rm d\theta} \, {\rm dp} \,\bar f_i[\varepsilon_i(\theta,p)]\,\cos \theta \,
\end{equation}
where $\varepsilon_i = \frac{p^2}{2} + 1 - m_i\cos \theta$.
Writing,
\begin{equation}
    \bar f_i[\varepsilon_i(\theta,p)]=\int {\rm d\epsilon}\, \bar f_i(\epsilon) \,\delta\left[\varepsilon_{i}(\theta,p)-\epsilon\right]\,.
\end{equation}
and substituting this in Eq. (\ref{eq:map1a}) and exchanging the order of integration, we obtain:
\begin{widetext}
\begin{equation}
\label{eq:map1}
m_{i+1} = 4\begin{cases} 
      \int {\rm d\epsilon}\, \frac{m_i\{2E[\kappa_i(\epsilon)]-K[\kappa_i(\epsilon)]\}}{\sqrt{|m_i|^3}} \bar f_i(\epsilon) & \text{if } \kappa_i(\epsilon) \leq 1 \\
      \int {\rm d\epsilon} \, \frac{m_i\{(1-2\kappa_i(\epsilon)) K[\kappa_i^{-1}(\epsilon)] + 2\kappa_i(\epsilon) E[\kappa_i^{-1}(\epsilon)]\} }{\sqrt{\kappa_i(\epsilon) |m_{i}|^3}}\,\bar f_{i}(\epsilon) & \text{if } \kappa_i(\epsilon) > 1 
   \end{cases}
\end{equation}
\end{widetext}
where the lower limit of the integral is $1 - |m_i|$. \textcolor{black}{The expression for 
$\kappa_i(\epsilon)$ is  
the same as $\kappa_0(\epsilon)$, with the index $0$ replaced by $i$.} The updated magnetization \(m_{i+1}\), will change the  isoenergy  trajectories, so that the particles  that were distributed according to the coarse-grained distribution  \(\bar f_i(\epsilon)\) will now redistribute over the new isoenergy contours defined by \(\varepsilon_{i+1}(\theta, p)\), resulting in a new coarse-grained distribution:
\begin{equation}
\label{eq:map2a}
\bar f_{i+1}(\epsilon) = \frac{\iint {\rm d}\theta \, {\rm d}p \, \bar f_i[\varepsilon_i(\theta,p)]\, \delta[\varepsilon_{i+1}(\theta,p) - \epsilon]}{\iint {\rm d}\theta \, {\rm d}p \, \delta[\varepsilon_{i+1}(\theta,p) - \epsilon]} \,.
\end{equation}
Using the properties of the delta function, the integral over momentum can be performed explicitly resulting in:
\begin{equation}
\label{eq:map2}
\bar f_{i+1}(\epsilon) = \frac{\int {\rm d}\theta \, \frac{\bar f_i(\epsilon - \delta m_i \cos \theta)}{\sqrt{2(\epsilon - 1 + m_{i+1} \cos \theta)}}}{g_{i+1}(\epsilon)} \,,
\end{equation}
where \(\delta m_i = m_i - m_{i+1}\) and the expression for $g_{i+1}(\epsilon)$ is the same as in Eq.~\ref{g0}, with the index $0$ replaced by $i+1$.  The evolution described by the map Eqs. (\ref{eq:map1}) and (\ref{eq:map2}) terminates when $m_\infty$ reaches a fixed point.  In all cases studied the ferromagnetic state always corresponds to a fixed point of the map with $m \ne 0$, while paramagnetic state is found to be a fixed point with  $m_\infty=0$~\cite{Kaneko}. 



To compensate for the cooling of the core produced by the mixing process we introduce a halo region which extends from the maximum energy of the core particles, which we shall denote as the Fermi energy \( \epsilon_F \), up to  the halo energy  \( \epsilon_h \). The maximum energy of evaporating particles \( \epsilon_h \) can be calculated using the canonical perturbation theory~\cite{Gluckstern:1994,Levin:2013}. The population of particles in the halo region is assumed to be uniform in the phase space, consistent with the Landau damping mechanism that progressively decreases the  oscillation of the mean-field potential leading to the final qSS.  We thus propose a \textcolor{black}{local mixing core-halo (LMCH) ansatz} for the particle distribution in the qSS:
\begin{equation}
\label{eq:qSS}
\bar f_{qSS}(\epsilon)=\bar f_{\infty}(\epsilon)\Theta(\epsilon_F-\epsilon)+\chi \Theta(\epsilon-\epsilon_F)\Theta(\epsilon_h-\epsilon)\,.
\end{equation}
The core region, described by the map Eqs. (\ref{eq:map1}) and (\ref{eq:map2}), is bounded by the Fermi energy. Above \( \epsilon_F \) the particles are uniformly distributed up to \( \epsilon_h \).  The values of \( \epsilon_F \) and the phase space density of the halo particles $\chi$ are calculated self-consistently to preserve the total energy of the system and the norm of the distribution function.  
The marginal distributions can be obtained as follows:
\begin{eqnarray}
\label{eq:marginal}
\begin{cases}
n_i(\theta)=\int {\rm dp} \, \bar f_i[\varepsilon_i(\theta,p)]
\\
n_i(p)=\int {\rm d\theta} \, \bar f_i[\varepsilon_i(\theta,p)]\,.
\end{cases}
\end{eqnarray}
An example of the marginal distribution functions calculated using the present theory are 
shown in the Fig.~\ref{figinitial}.  As can be seen from the figure, there is an excellent agreement with the results of MD simulations, while LB \textcolor{black}{and CH theories} predict qualitatively incorrect form of the distribution function.

The HMF model is known to exhibit a phase transition between ordered and disordered phases~\cite{Campa:2009}. The  transition has been  investigated using MD simulations and \textcolor{black}{LB and CH theories}.  It was shown that LB incorrectly predicts the order of the phase transition even for the simplest one level water-bag initial distribution of particles~\cite{Antoniazzi,Pakter:2011}. Very little is known about the non-equilibrium phase transitions of \textcolor{black}{continuous} initial particle distributions.  The non-linear stability analysis based on Vlasov equation can be used to predict that an initial paramagnetic state with concave velocity distribution functions will  become unstable below some energy threshold~\cite{Yamaguchi:2004}.  However, nothing is known analytically about the qSS to which the system will relax or even if the qSS will remain paramagnetic or will become magnetized.  Nothing is known about the relaxation of initially convex (population inverted ) distributions.

\begin{figure}[!h]
\includegraphics[scale=1.25,width=8.cm]{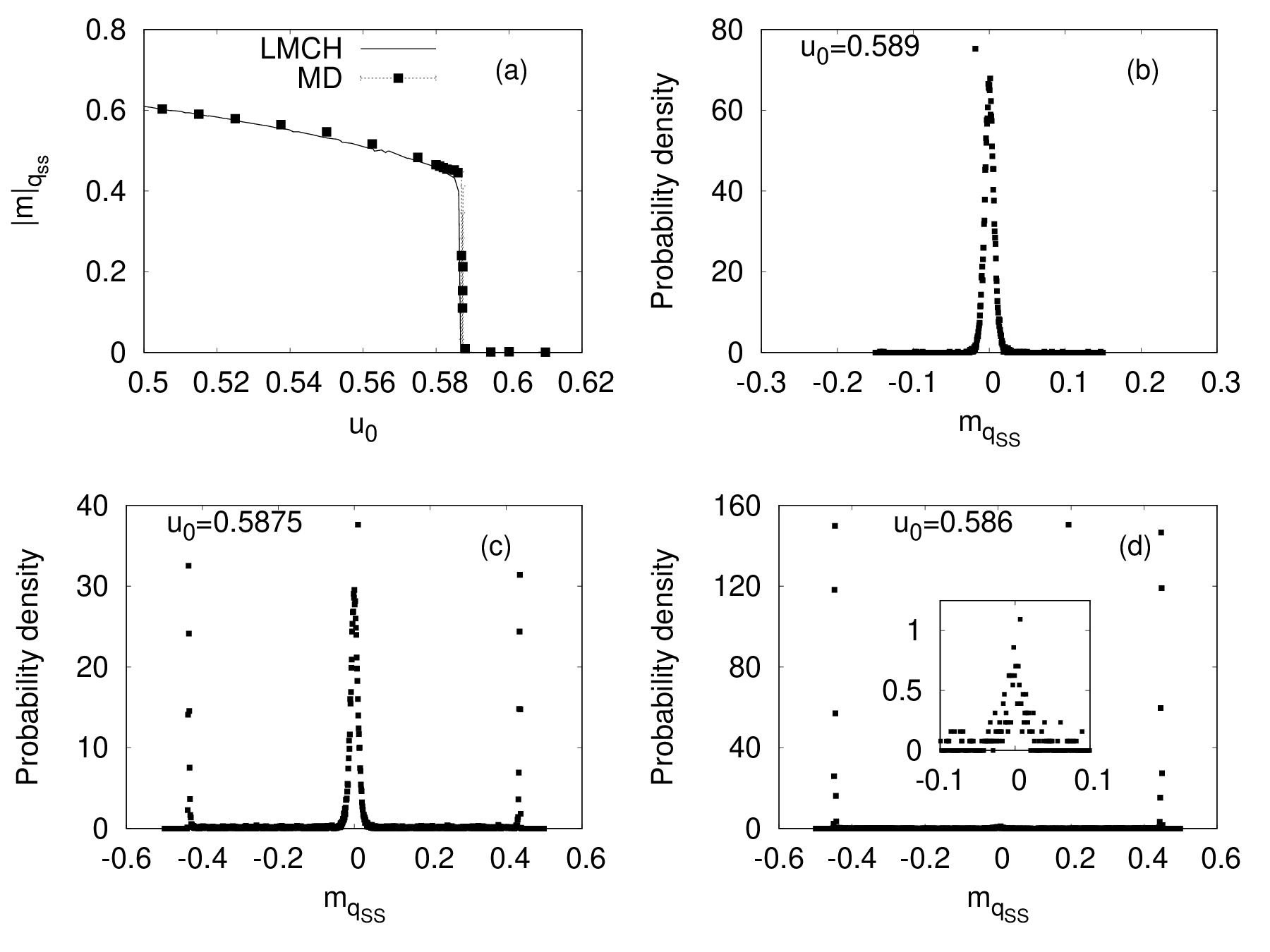}
\caption{The magnetization of the final qSS starting from initial water-bag particle distribution \( f_0(\theta,p)=\frac{1}{4\pi p_0}\Theta(p_0-|p|) \), as a function of the initial energy per particle  \( u_0 \). Panel (a) shows the modulus of magnetization \( |m|_{q_{SS}} \) in the final qSS, as a function of the initial energy. The solid lines represent theoretical predictions \textcolor{black}{of LMCH theory}, while symbols depict results of molecular dynamics (MD) simulations. Panels (b), (c), and (d) show histograms of magnetization for various  \( u_0 \) \textcolor{black}{constructed using $10^4$ different random realizations drawn from the initial distribution}.  The transition region is characterized by the coexistence of three peaks indicating a first order phase transition. \textcolor{black}{The inset in panel (d) is included to show the appearance of a small paramagnetic peak for this energy.} Away from the transition region, the the dispersion of magnetization is comparable to the size of the data points. The value of $m_{qSS}$ was computed after $t=5\times 10^3$ as an averaging over a time window of $\delta t=100$.
} 
\label{figwb}
\end{figure}

\begin{figure}[!h]
\includegraphics[scale=1.25,width=8.cm]{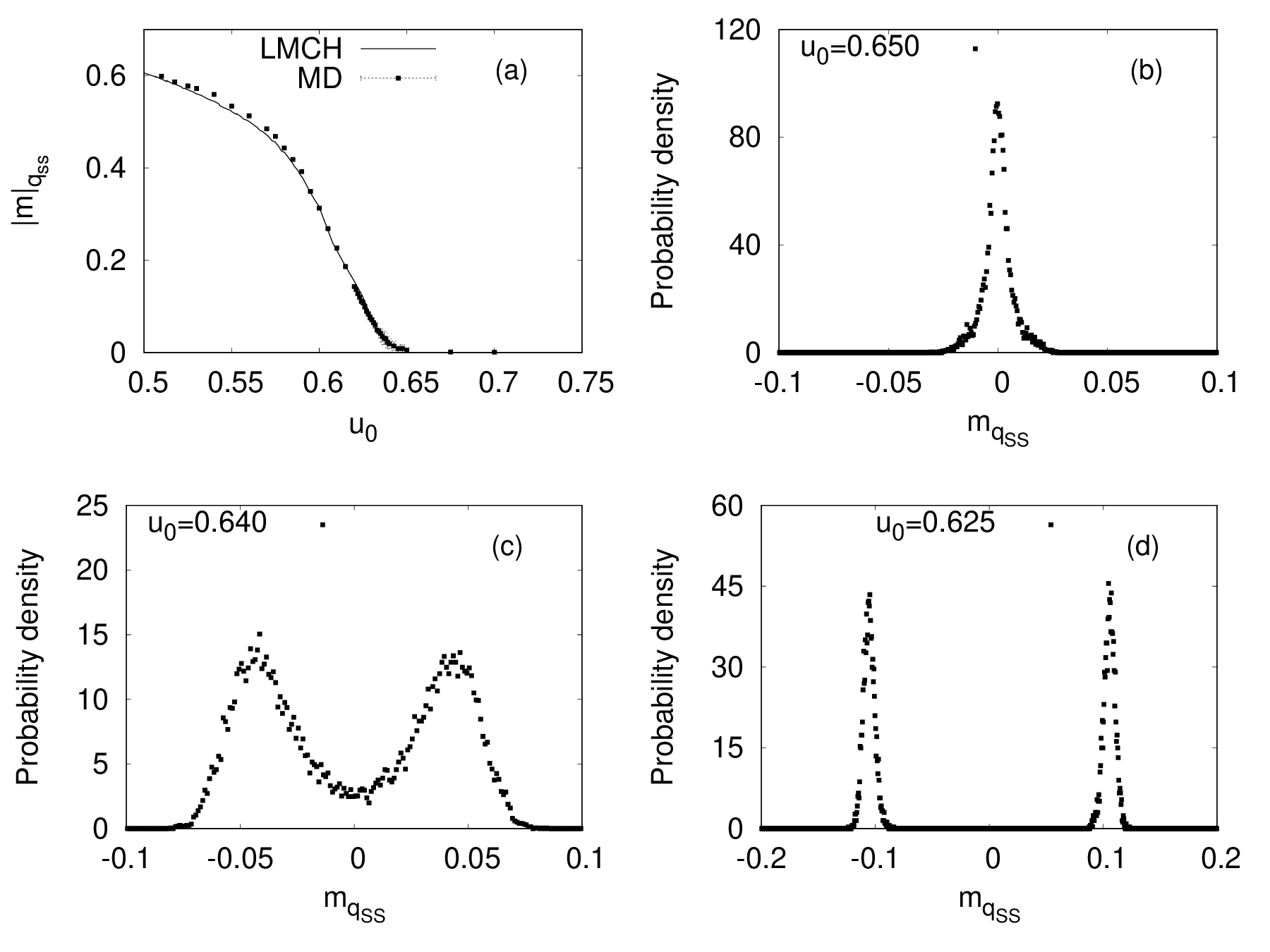}
\caption{Same as in Fig. \ref{figwb} but with initial concave particle distribution $f_0(\theta,p)=\frac{3}{8\pi p_0}\left[1-\left(\frac{p}{p_0}\right)^2\right]\Theta(p_0-|p|)$. 
} 
\label{figpm2}
\end{figure}

\begin{figure}[!h]
\includegraphics[scale=1.25,width=8.cm]{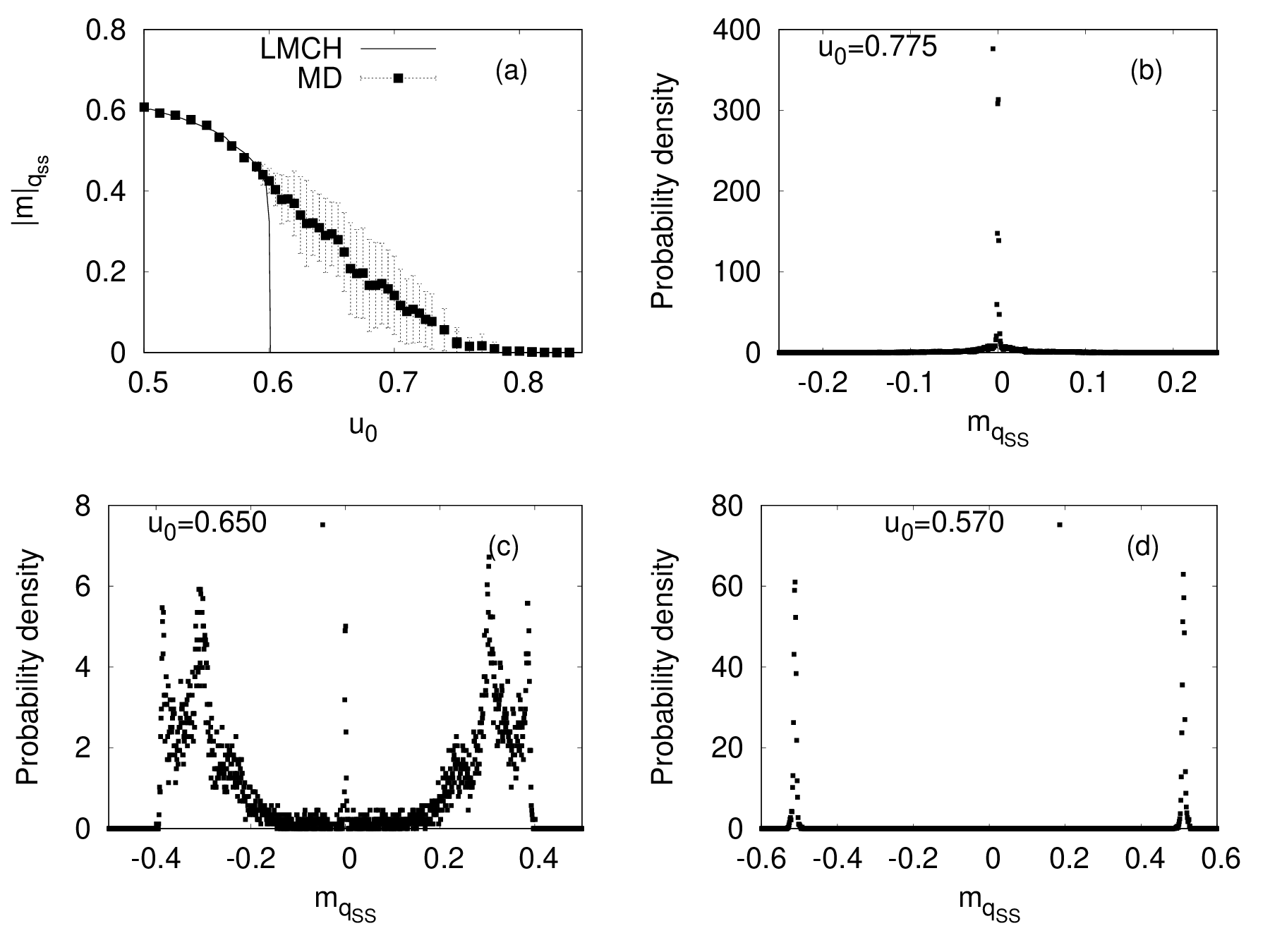}
\caption{Same as in Fig. \ref{figwb} but with initial convex (population inverted) particle distribution $f_0(\theta,p)=\frac{3}{4\pi p_0}\left(\frac{p}{p_0}\right)^2\Theta(p_0-|p|)$. The transition region is characterized by huge fluctuations of the order parameter, reflected by very large standard deviations in magnetization of the final qSS.  
} 
\label{figp2}
\end{figure}

To demonstrate the broad applicability of the LMCH theory,  we will examine the out-of-equilibrium phase transitions in the HMF model starting from an initial  paramagnetic state.   
We consider three distinct initial distributions: water-bag, concave, and convex.  In analogy with the equilibrium order-disorder phase transitions, we will define the transition to be first order if the magnetization histogram in the transition region exhibits three peaks, corresponding to the ``coexistence" of qSS states with $\pm m$ and $0$, magnetizations, see Figs. ~\ref{figwb}, \ref{figpm2}, and \ref{figp2}.  The histograms are constructed by tabulating the values of magnetization of the final qSS,  
\( m_{q_{SS}} \), to which the system evolves in MD simulation starting from an initial state drawn from the probability distribution $f_0(\theta,p)$.  In simulations we have used  \( N = 2 \times 10^5 \) particles.  The standard deviations and the histograms were calculated using  \( 10^4 \) random realizations of the initial conditions.  

\textcolor{black}{For the two initial distributions presented in  Figs.~\ref{figwb} and  \ref{figpm2}  one can use the non-linear stability analysis~\cite{Yamaguchi:2004,crawford1994,ogawa2014,Holm,Arnold} to predict the critical energy  at which initial paramagnetic water-bag and (concave) parabolic distributions will become unstable:  $u_0^*=7/12$ and $u_0^*=13/20$, respectively. These values of $u_0$ are close to when a coexistence of paramagnetic and ferromagnetic peaks is observed in MD simulations, see Figs. ~\ref{figwb}c, and \ref{figpm2}c. The LMCH theory also agrees with the results of the non-linear stability analysis, but in addition allows us to predict the particle distributions function and the magnetizations after the phase transition takes place, see  Figs. ~\ref{figwb}a, \ref{figpm2}a. }  
In the case of convex (population inverted) initial distributions, the non-linear stability theory can not be applied~\cite{Yamaguchi:2004}  and nothing is known about the phase transitions starting from such \textcolor{black}{initial states}.  In Fig.~\ref{figp2},  we show the predictions of \textcolor{black}{LMCH} applied to initial  population inverted parabolic distribution of particle velocities.  The simulations show that the transition region characterizing  convex distributions is extremely complex, with the probability measure of magnetized states described by a very broad distribution of magnetizations, see Fig.  \ref{figp2}c.  This is also reflected in the huge standard deviations that appear in the transition region of  Fig.  \ref{figp2}a.   Nevertheless, we see a ``coexistence" of three dominant peaks in the histogram, which allows us to denote the phase transition as  ``first order''.
The critical energy $u_0^*$ at which a paramagnetic $m=0$ peak first appears agrees with the location of the first order phase transition predicted by the theory, see Fig.  \ref{figp2}a.

\textcolor{black}{In summary, the theory presented in this paper enhances our understanding of collisionless relaxation in LR interactioning systems.  It integrates various approaches starting from LB 1967 pioneering work, into a new framework that is able to very accurately predict the final qSS to which a LR system will relax starting from an arbitrary initial particle distribution function. We find that despite the broken ergodicity, LR systems exhibit local mixing along the isoenergy curves. Our approach allows for a quantitative prediction of the final qSS to which a system will relax starting from an arbitrary initial particle distribution function. The paper also clarifies the nature of nonequilibrium phase transition in the HMF model. As the energy per particle in the initial distribution is decreased, we observe a first-order phase transition from a paramagnetic to a ferromagnetic state for both the WB and convex initial distributions, and a continuous transition in the case of the concave initial distribution. Notably, in the first-order cases, before the paramagnetic state fully disappears, we identify a coexistence region in which initial conditions drawn from exactly the same distribution can lead to either a paramagnetic or a ferromagnetic qSS, with different probabilities.    
This represents a first clear demonstration of the coexistence of different qSS in systems with LR interactions.  
In the future work we will explore applications of the present theory to self-gravitating systems and magnetically confined plasmas.}

\end{document}